\documentclass[]{iopart}

\usepackage{amssymb} 

\expandafter\let\csname equation*\endcsname\relax

\expandafter\let\csname endequation*\endcsname\relax

\usepackage{amsmath}

\usepackage{booktabs}

\usepackage{subfigure}

\usepackage{graphicx}% Include figure files

\graphicspath{{images/}}

\usepackage[colorlinks=true,allcolors=blue]{hyperref}

\usepackage{xcolor, soul}

\newcommand{\ket}[1]{\left| #1 \right>}

\newcommand{\round}[1]{\ensuremath{\lfloor#1\rceil}}

\begin{document}
	
\title[Generalized adiabatic approximation to the asymmetric quantum Rabi model]{Generalized adiabatic approximation to the asymmetric quantum Rabi model: conical intersections and geometric phases}

\author{Zi-Min Li$ ^{1} $, Devid Ferri$ ^{1} $, David Tilbrook$ ^{1} $ and Murray T. Batchelor$ ^{2,1,3} $}
\address{
	$ ^1 $ Department of Theoretical Physics, Research School of Physics, Australian National University, Canberra ACT, 2601, Australia\\
	$ ^2 $ Mathematical Sciences Institute, Australian National University, Canberra ACT, 2601, Australia\\
	$ ^3 $ Centre for Modern Physics, Chongqing University, Chongqing, 400044, People's Republic of China
}
\ead{murray.batchelor@anu.edu.au}

\date{\today}

\vspace{10pt}
%\begin{indented}
%	\item April 2021
%\end{indented}

\begin{abstract}
The asymmetric quantum Rabi model (AQRM), which describes the interaction between a quantum harmonic oscillator and a biased qubit, arises naturally in circuit quantum electrodynamic circuits and devices. The existence of hidden symmetry in the AQRM leads to a rich energy landscape of conical intersections (CIs) and thus to interesting topological properties. However, current approximations to the AQRM fail to reproduce these CIs correctly. To overcome these limitations we propose a generalized adiabatic approximation (GAA) to describe the energy spectrum of the AQRM. This is achieved by combining the perturbative adiabatic approximation and the exact exceptional solutions to the AQRM. The GAA provides substantial improvement to the existing approaches and pushes the limit of the perturbative treatment into non-perturbative regimes. As a preliminary example of the application of the GAA we calculate the geometric phases around CIs associated with the AQRM. 
\end{abstract}

\maketitle

\section{Introduction}

A two-level system coupled to a quantum harmonic oscillator is arguably the simplest and most ubiquitous interaction model in quantum physics. 
Such a system is usually referred to as the quantum Rabi model (QRM) \cite{Rabi_1936,Rabi_1937,Xie_2017} or the single mode spin-boson model. 
Due to the rapid development of quantum technologies, circuit quantum electrodynamic (cQED) systems are providing an increasingly powerful platform for the quantum simulation of light-matter interactions, initially at the level of a few quanta, but more recently with regard to many-body physics \cite{cQEDa,cQEDb,cQEDc}. 
Compared to the QRM, the asymmetric quantum Rabi model (AQRM), in which the Hamiltonian characterising the two-level system contains non-vanishing off diagonal entries, is more naturally adapted to cQED systems \cite{Niemczyk2010,Yoshihara_2016}. 
The AQRM has the advantage, for example, that it characterises cQED systems in which some of the experimental parameters may be tuned by the application of magnetic flux through a circuit loop.

While cQED experiments have managed to reach the ultra-strong and deep-strong coupling regimes 
\cite{Frisk2019,Forn2019,Blais2020}, the theoretical aspects are far from fully understood.
A major development from the theoretical viewpoint is the consideration of integrability and solvability of the QRM and its relatives. 
The current analytic solutions determine the energy spectrum by the zeros of transcendental functions \cite{Braak_2011,Chen2012,Zhong_2014,Maciejewski_2014}. 
In particular, there is no simple closed-form solution to the general spectrum of the AQRM.
Rather, only some isolated solutions known as exceptional solutions \cite{Li_2015,Wakayama_2017,Kimoto_2020} are found to be determined from polynomials.

A number of approximations have thus been proposed to deal with the AQRM \cite{Li2021}. 
Examples include the generalized rotating-wave approximation \cite{Zhang_2013,Mao_2018,Xie2020}, 
the adiabatic approximation (AA) \cite{Irish_2005,Semple_2017,Ashhab_2020} and van Vleck perturbation theory \cite{Hausinger_2010}.
In the limit where the qubit frequency is much smaller than the field frequency, 
the AA gives the simplest expressions for the eigenstates and corresponding eigenvalues. 
Accordingly, the AA has been adapted widely by theorists and experimentalists alike.

An interesting feature of the AQRM is that the energy landscape contains conical intersections (CIs) \cite{Batchelor2016} around which 
non-vanishing geometric phases \cite{Berry1984} would be expected.  
Geometric phases and related topological phenomena have been identified and studied in many areas of physics, 
most notably in condensed-matter physics and optics \cite{SW1989,Bohm2003,Cohen_2019,Larson2020}. 
The geometric phase also offers opportunities for applications in quantum information and computation \cite{Cohen_2019}. 
There have been a number of experimental developments of relevance to two-level systems.
For example, the controlled accumulation of a geometric Berry phase in a superconducting charge qubit, 
manipulating the qubit geometrically using microwave radiation, and the associated accumulated phase has been demonstrated \cite{Leek_2007}.
Further to this experiment, a vacuum-induced Berry phase has been measured 
in the artificial atom interacting with the single microwave cavity mode \cite{Gasparinetti_2016}. 
The artificial atom acquires a geometric phase determined by the path traced out in the combined Hilbert space of the atom 
and the quantum field as the phase of the interaction is varied.
Measurements of the Berry phase in a superconducting charge pump have also been performed showing the 
dependence of both dynamic and geometric effects on the superconducting phase bias across the pump \cite{Mottonen_2008}. 
Measurements of a superconducting phase qubit have demonstrated a contribution from the second 
excited state to the two-level geometric phase in a weakly anharmonic but strongly driven two-level system \cite{Berger_2012}.

The geometric Berry phase \cite{Berry1984} in the QRM has been discussed by several authors \cite{Fuentes_Guridi_2002,Larson_2012,Larson2020}. 
In that work the geometric phase is induced by a unitary transformation \cite{Deng_2013,Wang_2015,Mao_2015,Calder_n_2016,Wang2019}. 
Here we will consider geometric phases associated with CIs in the AQRM.
In general it should be expected that the geometric phase is a multiple of $\pi$ \cite{Berry1984,SW1989}.
In particular, the Herzberg/Longuet-Higgins theorem \cite{HLH_1963} 
implies that a two-level system with a real symmetric Hamiltonian always acquires a non-trival $\pi$ phase  around a CI. 
Analysis of these phases in the AQRM does not seem tractable using the analytic solutions. 
Moreover, none of the existing approximations are capable of recovering the CIs in the AQRM. 
The correct location of the CIs is essential for the calculation of topological properties.

In this paper, we propose an approximation which exactly recovers the CIs in the AQRM. 
This is achieved based on a combination of the AA \cite{Irish_2005,Hausinger_2010}  
and the exact exceptional solutions \cite{Li_2015,Kimoto_2020} for the AQRM. 
This approach has been applied successfully to the QRM and is called the generalized adiabatic approximation (GAA) \cite{Li2021GAA}.

The paper is organized as follows. 
In Sec.~\ref{SectionCI} we describe the model Hamiltonian and explain the CIs in the AQRM. 
We propose the GAA for the AQRM, which 
correctly describe the CIs, in Sec.~\ref{SectionGAA}. 
As an example of application, geometric phases around the CIs are calculated in Sec.~\ref{SectionGP}. 
Further discussion and concluding remarks are given in Sec.~\ref{SectionConclusions}.

\section{Conical intersections in the AQRM}\label{SectionCI}

The AQRM is defined by the Hamiltonian ($\hbar=1$)
\begin{equation}\label{AQRM}
	H = \dfrac{\Delta}{2}\sigma_z + \omega a^\dagger a + g\sigma_x \left( a^\dagger + a \right) + \dfrac{\epsilon}{2}\sigma_x,
\end{equation}
where $\sigma_x$ and $\sigma_z$ are Pauli matrices for a two-level system with level splitting $\Delta$ and bias $\epsilon$. 
The single mode bosonic field is described by the creation and annihilation operators $a^\dagger$ and $a$, and frequency $\omega$. 
The interaction between the two systems is via the coupling $g$. 

When $\epsilon=0$, Eq.~(\ref{AQRM}) reduces to the standard quantum Rabi model, which conserves the parity of excitation numbers, corresponding to $\mathbb{Z}_2$ symmetry \cite{Braak_2019}. 
It follows that the Hamiltonian can be decomposed into two blocks with definite parity. 
Energy levels from different parity sectors are allowed to cross.

In general, non-zero values of $\epsilon$ break this symmetry and the level crossings are avoided, leading to {Dirac-like cones} in the energy spectrum. 
In some special cases, where $\epsilon$ is a multiple of the field frequency $\omega$, the level crossings reappear, without any apparent symmetry. 
This phenomenon is referred to as {parity-like} hidden symmetry \cite{Wakayama_2017,Ashhab_2020}, 
which gives rise to {additional} CIs and rich topological properties. 
Operators responsible for the hidden $\mathbb{Z}_2$ symmetry in the AQRM have recently been found \cite{Mangazeev_2021,RBW2021}.
More generally, this hidden symmetry is seen to be universal in asymmetric light-matter interaction models \cite{Li2021a}, with corresponding 
hidden symmetry operators found for a number of asymmetric generalizations of the QRM \cite{Lu_a,Lu_b}.

\begin{figure}[t]
	\centering
	\includegraphics[width=.6\linewidth]{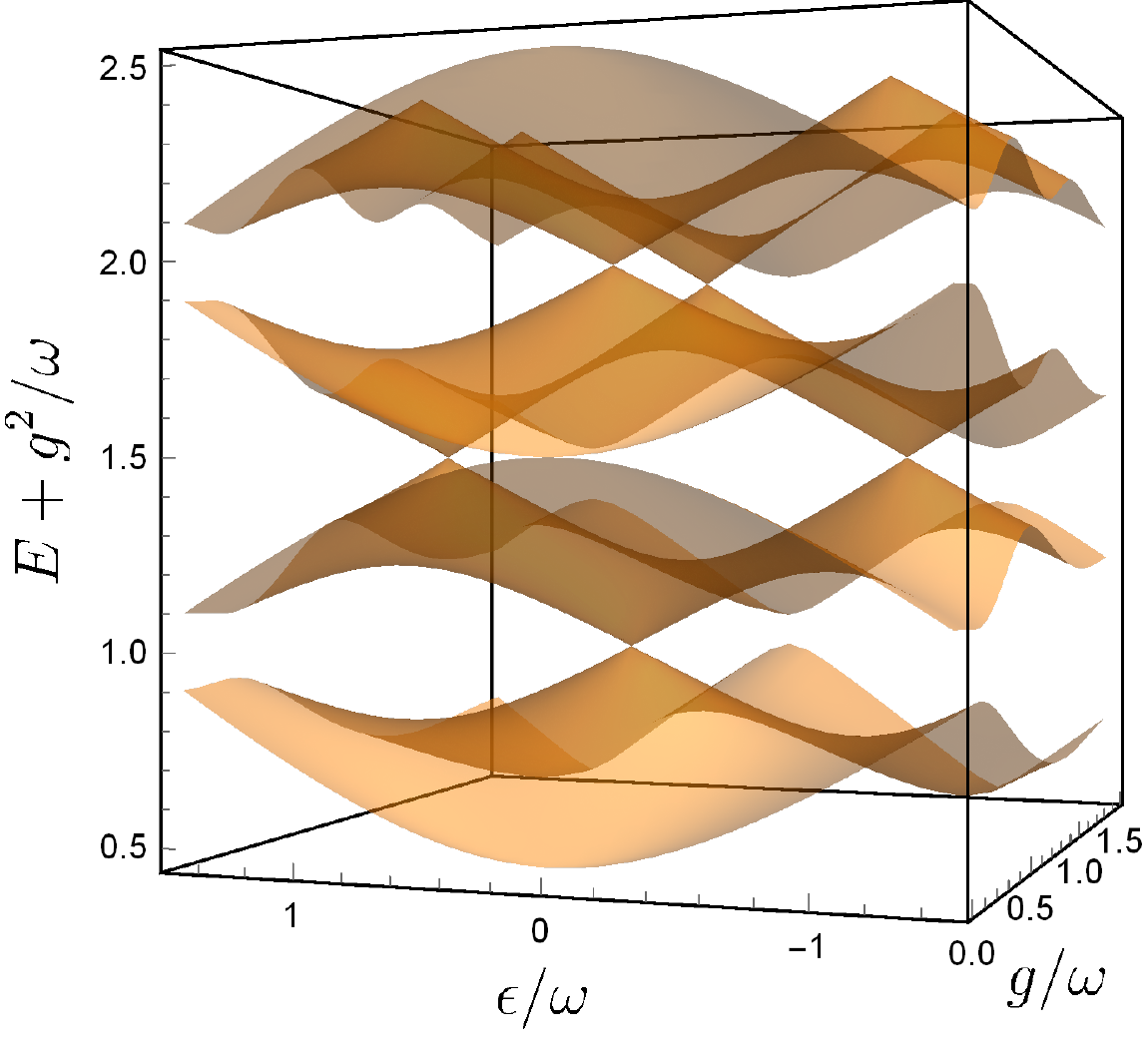}
	\caption{Energy spectrum of the AQRM with respect to the parameters $g$ and $\epsilon$. Other parameter values are $\omega=1$ and $\Delta=1$. Conical intersections emerge only when $\epsilon/\omega$ takes integer values. For clarity, the lowest two levels are not shown and the energies are rescaled with $E+g^2/\omega$. In general the coordinates of the CI points are integer values of $\epsilon/\omega$ and half-integer values of $E+g^2/\omega$. }
	\label{EnergyLandscape}
\end{figure}

A region of the energy landscape of the AQRM, including several of the CIs, is shown in Fig.~\ref{EnergyLandscape}.

~

\section{Generalized adiabatic approximation to the AQRM}\label{SectionGAA}

\subsection{Adiabatic approximation to the AQRM}

Although the exact solution of the AQRM is known \cite{Braak_2011,Chen2012,Zhong_2014,Maciejewski_2014}, 
it seems unlikely that it can be used to calculate the geometric properties of the AQRM analytically.
With this particular calculation in mind, we resort to the useful adiabatic approximation, which is able to produce CIs under some conditions.

Firstly we consider the AQRM in the limit $\Delta=0$.
In this case, with some reorganization, Eq.~(\ref{AQRM}) becomes
\begin{equation}\label{Hdo}
			H^\text{do} = \omega \left[\left(a^\dagger+ \dfrac{g}{\omega}\sigma_x\right)\left(a + \dfrac{g}{\omega}\sigma_x\right)\right] - \dfrac{g^2}{\omega} +\dfrac{\epsilon}{2} \sigma_x,
\end{equation}
which describes spin-dependent displaced harmonic oscillators with energies shifted by the bias term \cite{Irish_2005,Li2021a}.

Since the two oscillators are independent of each other and no coupling exists, Eq.~(\ref{Hdo}) can be readily solved. 
The eigenstates and eigenvalues are
\begin{equation}\label{DOEigenstates}
	\begin{aligned}
		&\psi^\text{do}_{n,\pm} =  |n_\pm,\pm\rangle =  | n_\pm \rangle \otimes |\pm\rangle , \\
		&E_{n,\pm}^\text{do} = n\omega - \dfrac{g^2}{\omega} \pm \dfrac{\epsilon}{2},
	\end{aligned}
\end{equation}
in which $|n_\pm\rangle = \exp[\mp {g(a^\dagger - a)}/{\omega}]|n\rangle$ are displaced Fock states or generalized coherent states \cite{Philbin_2014}, and $|\pm\rangle$ are the eigenstates of $\sigma_x$.

\begin{figure}[t]
	\centering
	\subfigure[]{
		\includegraphics[width=.31\linewidth]{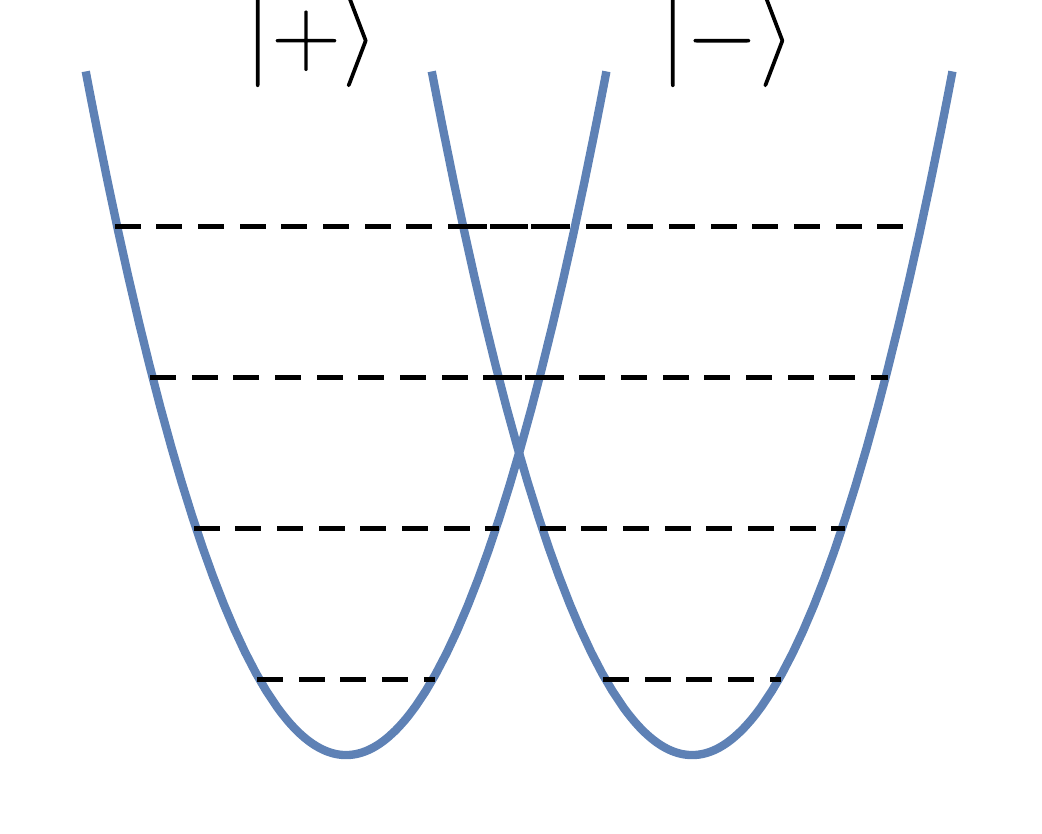}
		\label{potentiala}
	}
	\subfigure[]{
		\includegraphics[width=.31\linewidth]{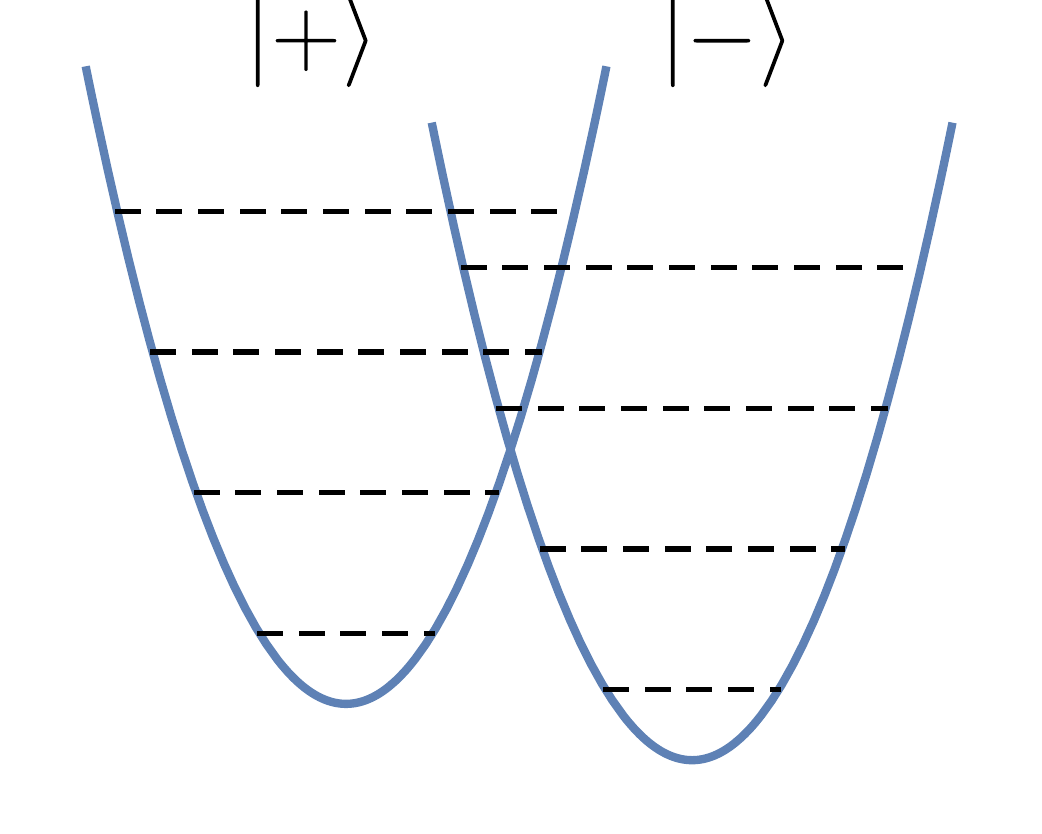}
		\label{potentialb}
	}
	\subfigure[]{
		\includegraphics[width=.31\linewidth]{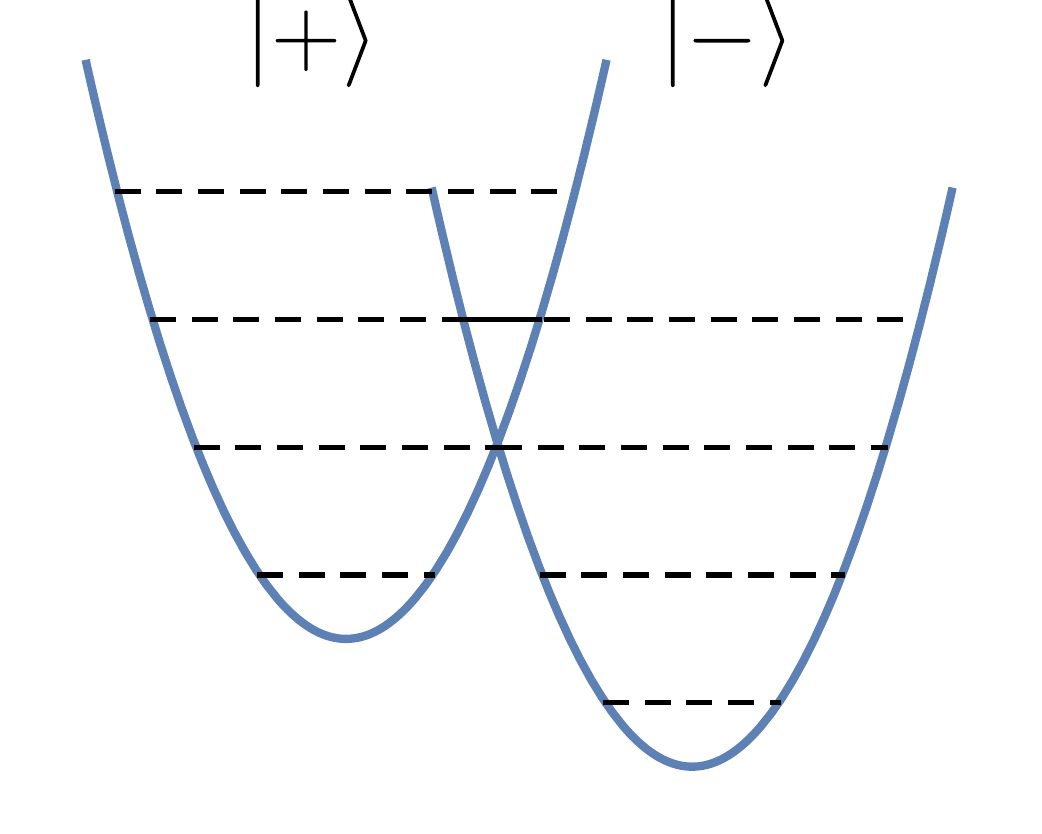}
		\label{potentialc}
	}
	\caption{Graphical representation of the displaced oscillator picture. 
		(a) When $\epsilon=0$, the AQRM reduces to the standard QRM and the oscillators are symmetric. Therefore, all energy levels are doubly degenerate. 
		(b) When $\epsilon\ne 0$, energies of the displaced oscillators are shifted by the amount of $ \pm{\epsilon}/{2} $, whose sign depends on the corresponding qubit states $ \ket{\pm}  $. In general, energy levels are not degenerate. 
		(c) In a special case where $\epsilon$ is a multiple of $\omega$, most energy levels are doubly degenerate again while the lowest few levels stay unpaired. 
 		} 
	\label{DOAQRM}
\end{figure}

Without loss of generality, $\epsilon$ {can be} assumed to be non-negative since the AQRM is symmetric with respect to $\epsilon$. 
When $\epsilon/\omega =0$, eigenstates $ \ket{n_+,+} $ and $ \ket{n_-,-} $ are degenerate, as shown in Fig.~\ref{potentiala}, 
which corresponds to the symmetric oscillators in the original QRM. 
Nonzero values of $\epsilon$ shift the energies of the oscillators and, as shown in Fig.~\ref{potentialb}, 
the degeneracies usually vanish. 
However, in the special case where $ \epsilon/\omega $ takes an integer value $ l $, the shifted levels $ \ket{n_+,+} $ and $ \ket{(n+l)_-,-} $ become degenerate, while the lowest $ l $ levels remain unpaired \cite{Li2021a}.\footnote{Excellent approximation of the unpaired levels has been reported recently in the further study of the hidden symmetry of the AQRM \cite{RW2021}.}
The example of $ \epsilon/\omega=1 $ is shown in Fig.~\ref{potentialc}.

Now consider the case when $\Delta$ is non-zero but $\Delta/\omega\ll 1$, corresponding physically to 
weak tunnelling between the two displaced oscillators. 
We further assume that $\epsilon/\omega-l\ll 1$, where $l = \round{\epsilon/\omega}$ is the integer closest to $\epsilon/\omega$ \cite{Li2021a}. 
Therefore, the weak tunnelling induced by the $\Delta$ term only couples the state pairs $|n_+,+\rangle$ and $|(n+l)_-,-\rangle$, 
whereas the lowest $l$ levels associated with the qubit state $|-\rangle$ remain uncoupled. 
This is known as the adiabatic approximation (AA) \cite{Irish_2005,Hausinger_2010, Semple_2017}, where ``adiabatic" implies that the qubit transition $\Delta$ is slow compared to the field frequency $\omega$. 
As a consequence, Eq.~(\ref{AQRM}) is block-diagonal in the basis $\left\{|n_+,+\rangle,|(n+l)_-,-\rangle\right\}$, 
with the $n$th $2\times 2$ matrix block given by
\begin{equation}\label{AABlockSigmaForm}
	H_n^\text{AA} = \left(n+\dfrac{l}{2}\right)\omega  - \dfrac{g^2}{\omega} + \dfrac{\epsilon- l \omega}{2} \sigma_x^{(n)} + \dfrac{\Omega_{nl}^\text{AA}}{2}\sigma_z^{(n)}. 
\end{equation}
The Pauli matrices become
\begin{equation}\label{AASigmas}
	\begin{aligned}
		&\sigma_x^{(n)} = |n_+,+\rangle \langle n_+, + | - | (n+l)_-, -\rangle \langle (n+l)_-,- |, \\
		&\sigma_z^{(n)} = |n_+,+\rangle \langle (n+l)_-, - | + | (n+l)_-, -\rangle \langle n_+,+ |,
	\end{aligned}
\end{equation}
and the off-diagonal tunnelling terms are
\begin{equation}\label{AAoverlap}
	\begin{aligned}
		\Omega_{nl}^\text{AA} = \Delta \exp\left[{-\dfrac{2g^2}{\omega^2}}\right]\left(-\dfrac{2g}{\omega}\right)^l\sqrt{\dfrac{n!}{(n+l)!}}L_n^l\left(\dfrac{4g^2}{\omega^2}\right),
	\end{aligned}
\end{equation}
which are given in terms of the generalized Laguerre polynomials $L_n^l$. 
It is then straightforward to obtain the eigenstates for the coupled levels as
\begin{equation}\label{AAEigenstates}
	\begin{aligned}
		&\psi^\text{AA}_{n,+} =  \cos\dfrac{\theta_n}{2}| n_+, + \rangle + \sin\dfrac{\theta_n}{2}| (n+l)_-, - \rangle, \\
		&\psi^\text{AA}_{n,-} =  -\sin\dfrac{\theta_n}{2}| n_+, + \rangle + \cos\dfrac{\theta_n}{2}| (n+l)_-, - \rangle, \\
	\end{aligned}
\end{equation}
where $\theta_n$ is determined by
\begin{equation}\label{AATheta}
	\tan\theta_n = \dfrac{\Omega_{nl}^\text{AA}}{\epsilon-l\omega}. 
\end{equation}
The corresponding eigenvalues are given as
\begin{equation}\label{AAEigenvalues}
	E_{n,\pm}^\text{AA} = \left(n+\dfrac{l}{2}\right)\omega - \dfrac{g^2}{\omega} \pm \dfrac{1}{2}\sqrt{\left(\Omega_{nl}^\text{AA}\right)^2 + \left(\epsilon-l\omega\right)^2}, 
\end{equation}
where index $ n $ is non-negative. 
The lowest $ l $ levels are unpaired and they are still described by the displaced oscillator in Eq.~(\ref{DOEigenstates}), 
i.e. $ E_m^\text{AA} = m\omega -g^2/\omega -\epsilon/2 $ with $ m = -l,...,-1 $. 

Up to this point, we have obtained simple expressions for the eigenvalues and eigenstates of the AQRM under the conditions $\Delta/\omega\ll 1$ and $\epsilon/\omega-l\ll 1$. 
These expressions are as simple as those obtained via the celebrated rotating-wave approximation. 
We aim to maintain this simplicity throughout this work.

\begin{figure}[t]
	\subfigure{	
		\includegraphics[width=.31\linewidth]{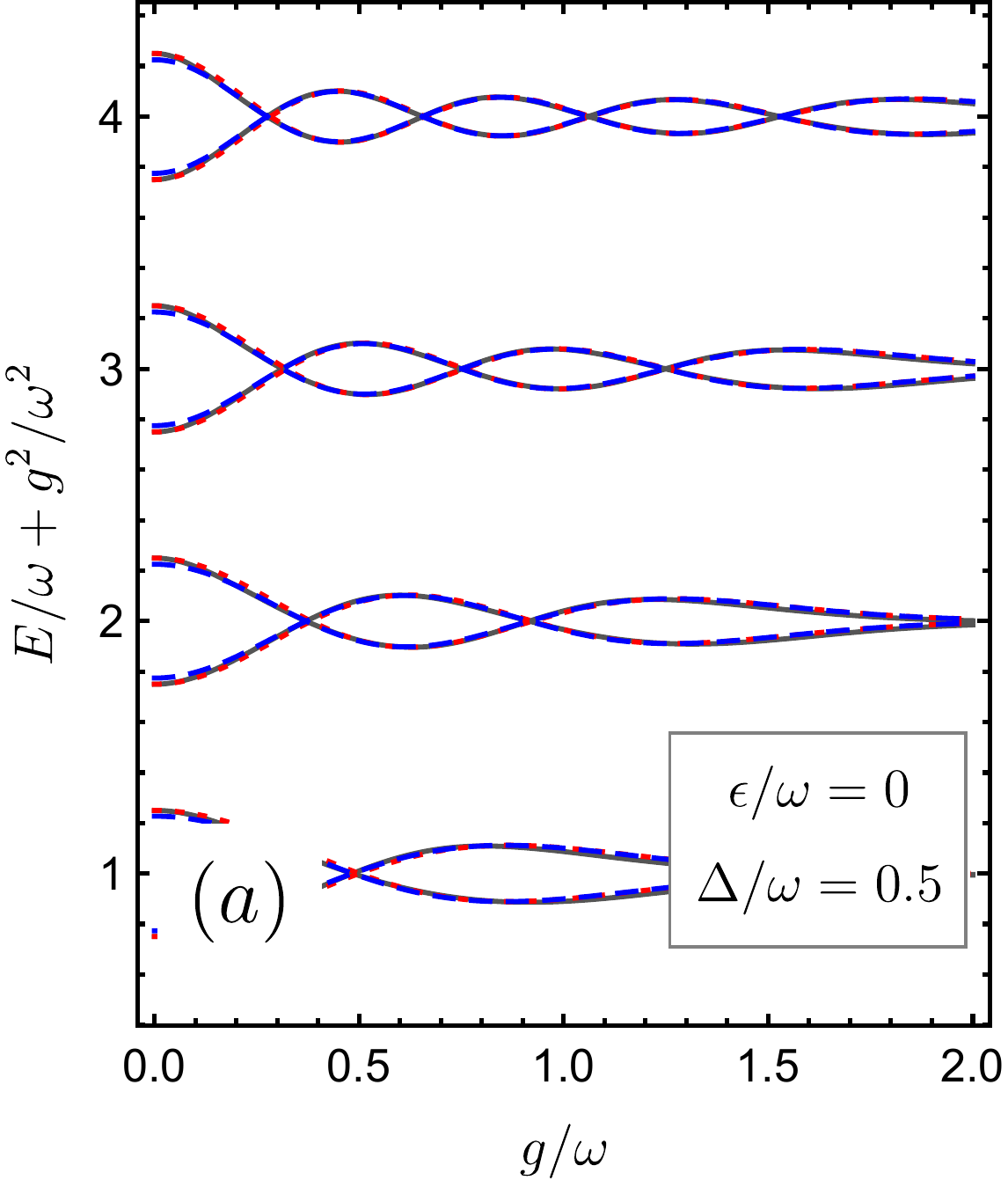}
		\label{spectra1}
	}
	\subfigure{	
		\includegraphics[width=.31\linewidth]{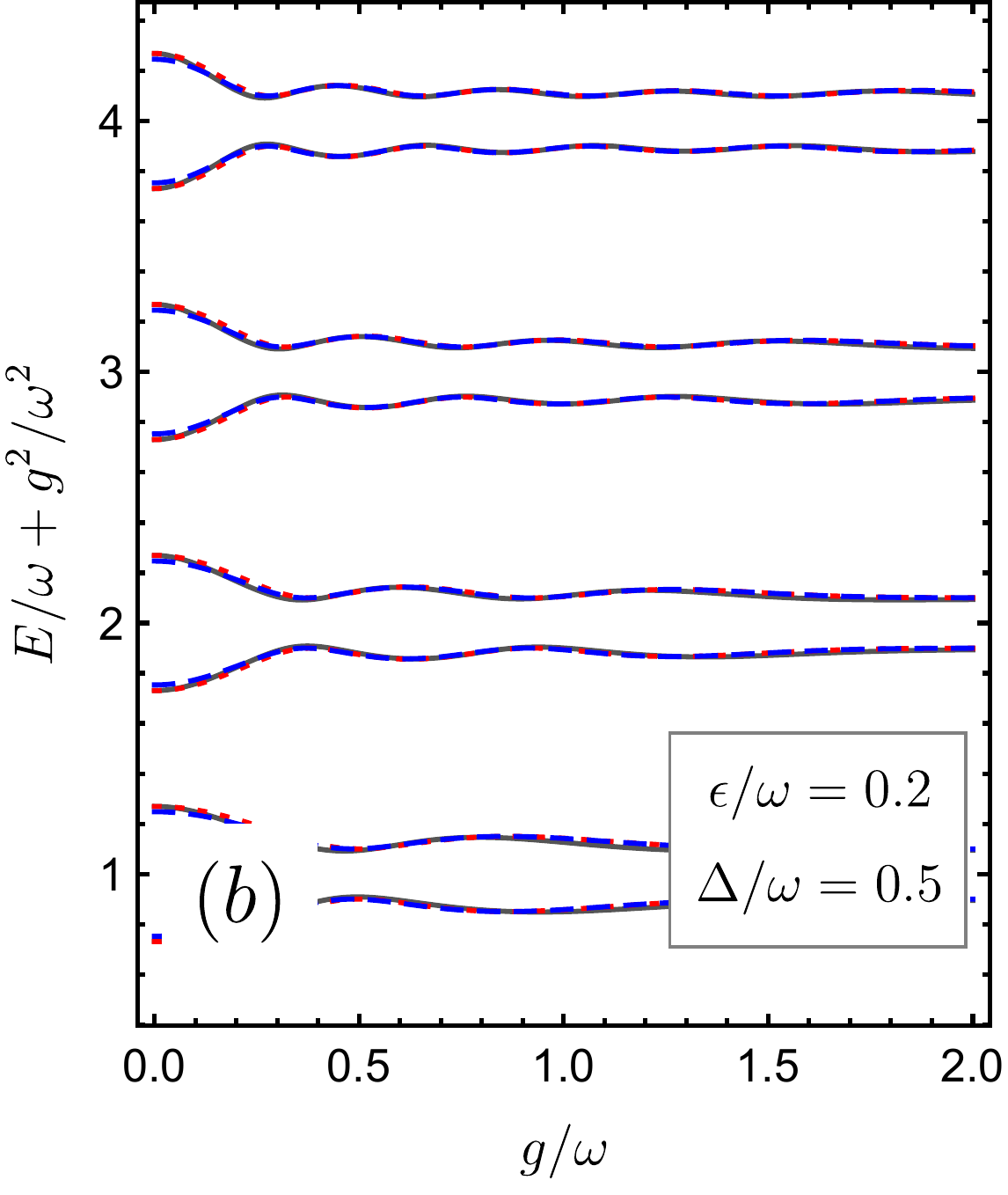}
		\label{spectra2}
	}
	\subfigure{	
		\includegraphics[width=.31\linewidth]{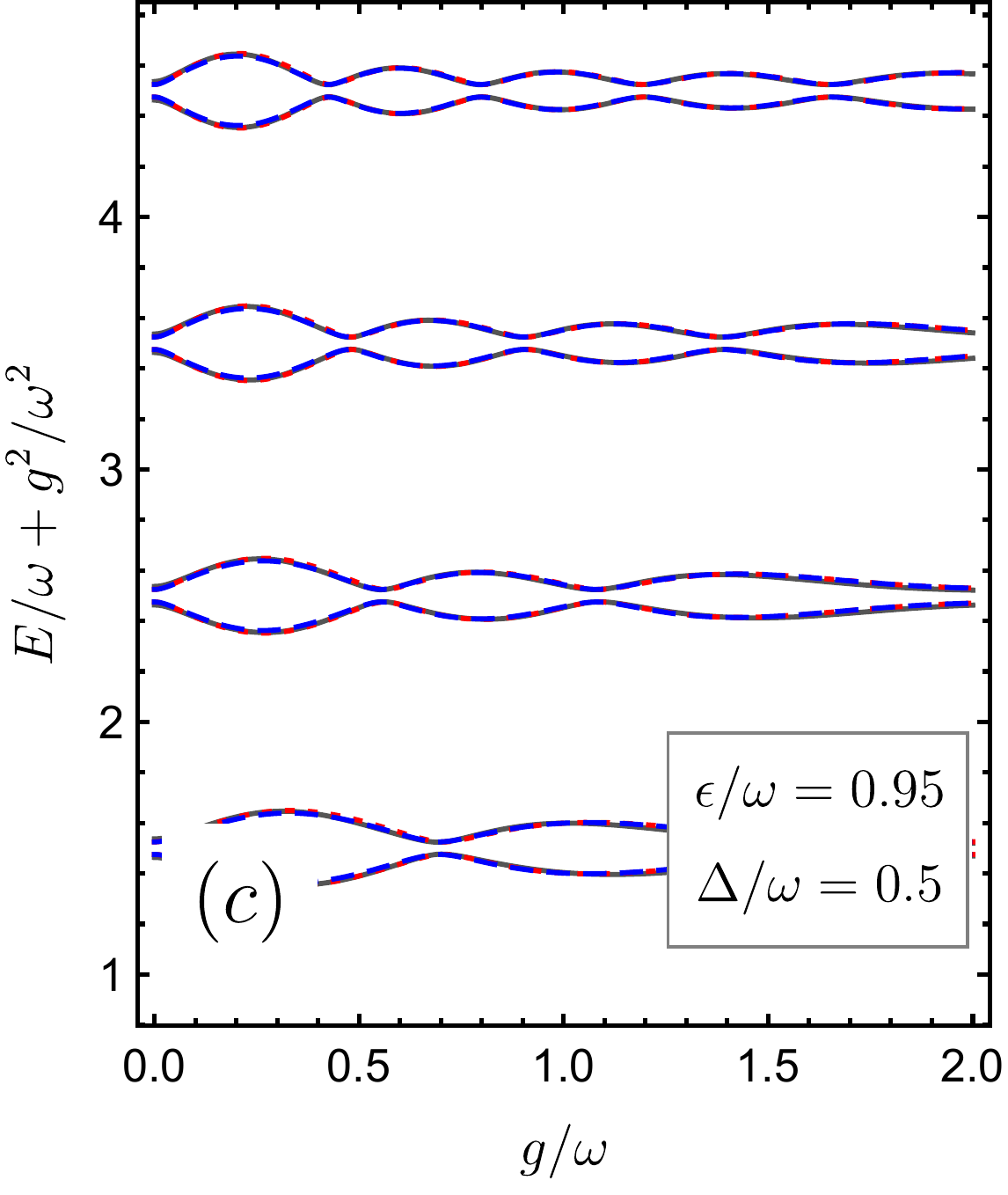}
		\label{spectra3}
	}
	\subfigure{	
		\includegraphics[width=.31\linewidth]{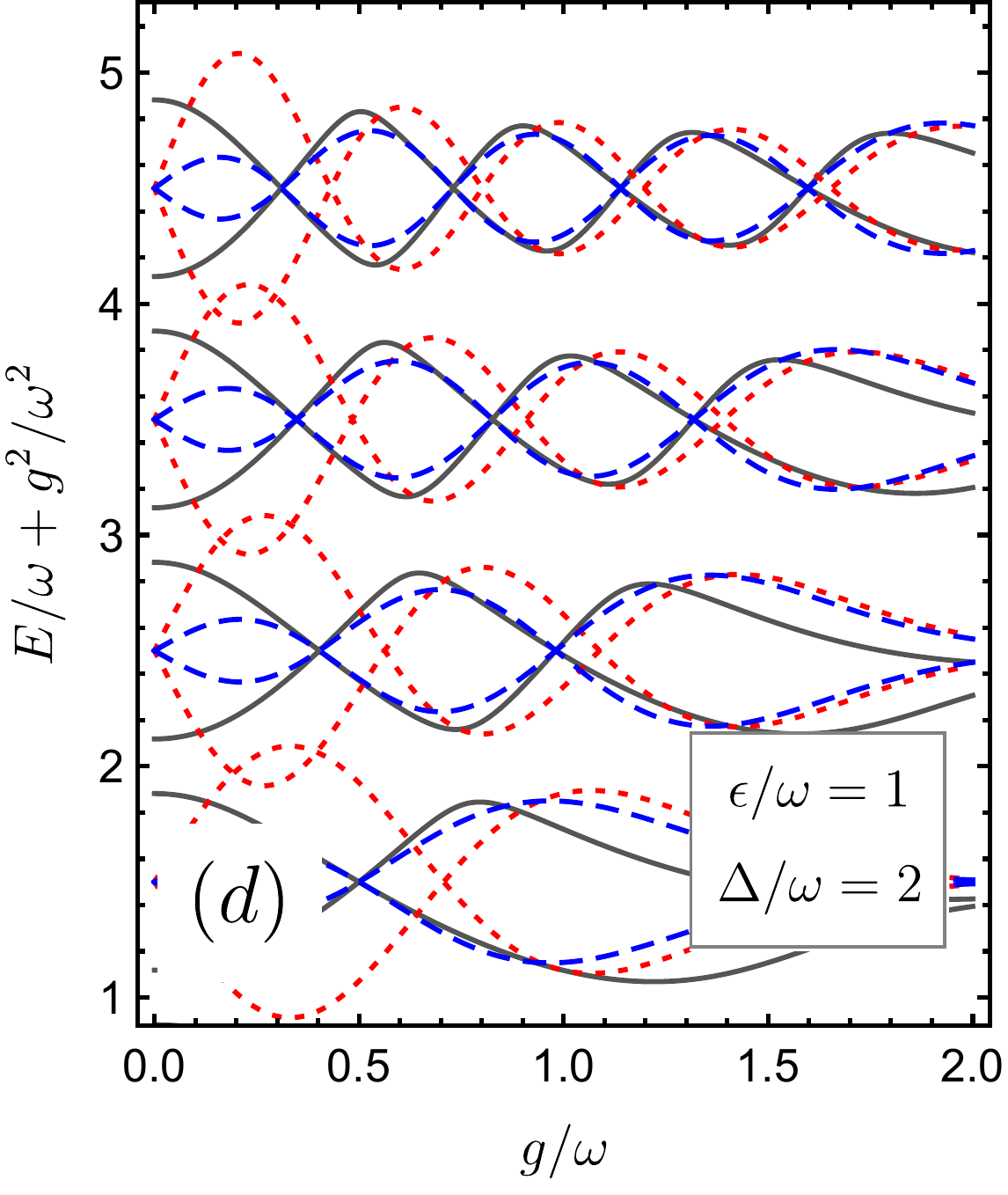}
		\label{spectra4}
	}
	\subfigure{	
		\includegraphics[width=.31\linewidth]{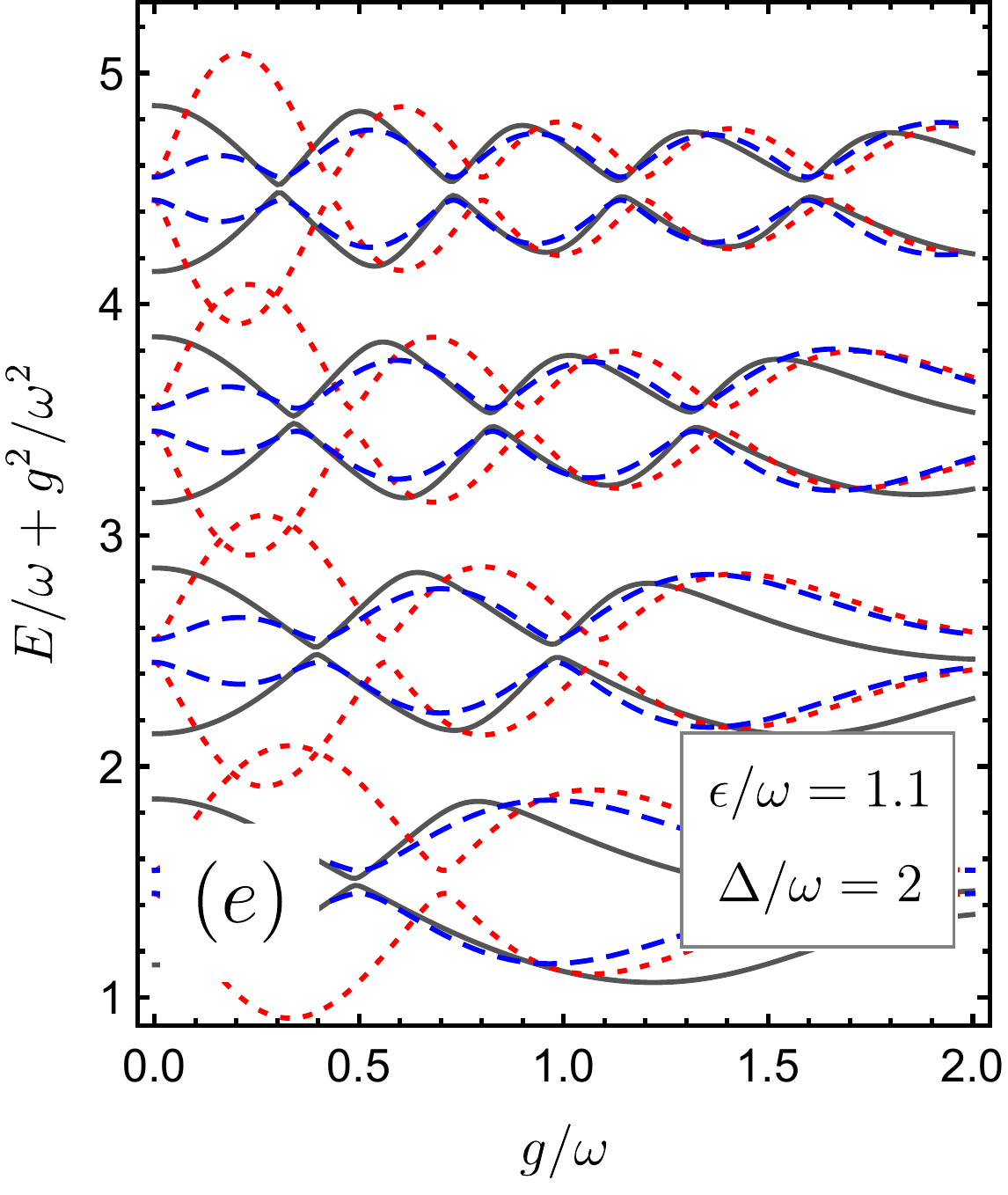}
		\label{spectra5}
	}
	\subfigure{	
		\includegraphics[width=.31\linewidth]{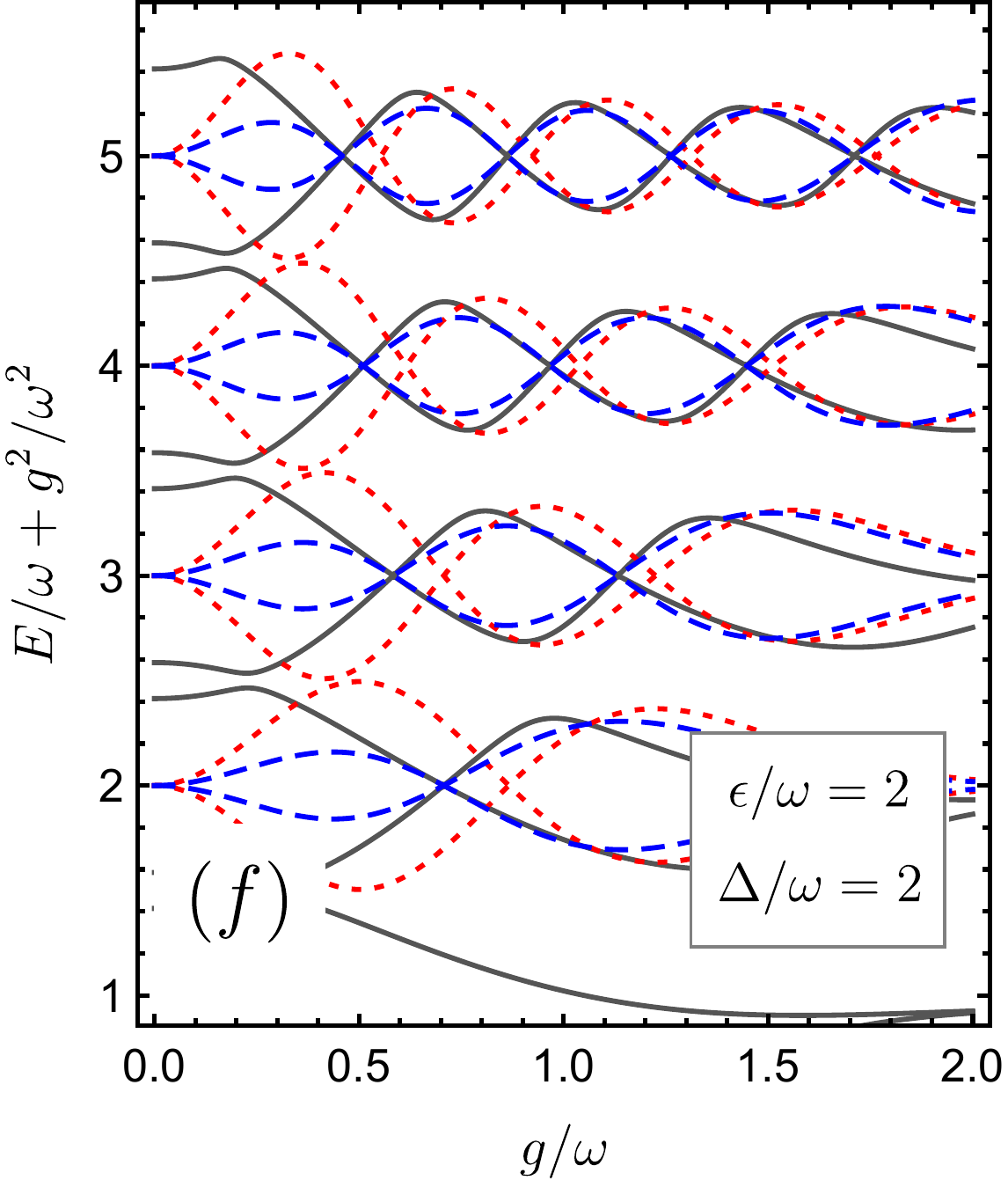}
		\label{spectra6}
	}
	\begin{center}
		\subfigure{ \includegraphics[width=0.5\linewidth]{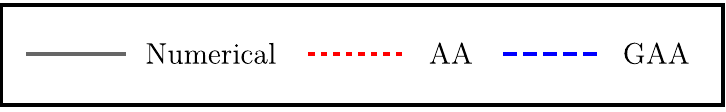}}
	\end{center}
	\caption{Generalized adiabatic approximation to the AQRM with various values of $\epsilon/\omega$ and $\Delta/\omega$. 
		(a)-(c) The parameter regimes where the AA agrees well with the exact results.  
		(d)-(f) The parameter regimes where the AA breaks down by predicting wrong crossing points and introducing unphysical crossings. The GAA solve these two problems nicely and substantially improve the overall performance. 
		For clarity, the lowest levels without crossings are not shown.  }
	\label{GAAE}
\end{figure}

The eigenvalues predicted by Eq.~(\ref{AAEigenvalues}) successfully approximate the overall ``ripple'' structure of the AQRM spectrum. 
Beyond the precondition $\Delta/\omega\ll 1$, these results also work reasonably well as long as $\Delta/\omega<1$. 
This can be confirmed from the comparison displayed in Fig.~\ref{GAAE}(a)-(c), where the AA results are denoted by red dotted lines. 
In these examples, $\Delta/\omega=0.5$ and the overall fittings look perfectly well.

Another important advantage of the AA is that energy level crossings only exist when $\epsilon/\omega$ strictly takes integer values, 
which is the key feature of the CIs in the AQRM \cite{Batchelor2016}. 
To demonstrate this point, we consider the source of level crossings from Eq.~(\ref{AAEigenvalues}). 
The energy gap between $ n $th level pair is given as
\begin{equation}
	\delta E_{n}^\text{AA} = E_{n,+}^\text{AA} - E_{n,-}^\text{AA} = \sqrt{\left(\Omega_{nl}^\text{AA}\right)^2 + \left(\epsilon-l\omega\right)^2}, 
\end{equation}
which is zero only when $ \Omega_{nl}^\text{AA} =0 $ and $ \epsilon/\omega $ is an integer. 

However, there are two shortcomings in the approximation. 
Firstly, beyond the condition $\Delta/\omega<1$, unphysical crossings appear in the spectrum predicted by the AA.
This results directly from the failure of the assumptions of the AA.
In the displaced oscillator picture, this means the coupling is strong enough and the tunnelling processes 
between non-degenerate states can no longer be neglected, with the perturbative treatment invalid.

Another problem with the AA is that the level crossings determined by Eq.~(\ref{AAEigenvalues}) deviate from the exact results.
This deviation can be understood from the energy expression Eq.~(\ref{AAEigenvalues}). 
Since CIs only occur when $\epsilon$ is an integer, the locations of crossing points are determined by the zeros of Laguerre polynomials in $ \Omega_{nl} $. 
The arguments of $ L_n^l $ do not contain any information about $\Delta$, meaning that the crossing points predicted by the 
AA are independent of $\Delta$, which is not true in the exact results. 
In fact, the tunnelling between non-degenerate levels is never strictly zero when $\Delta\ne0$, i.e. 
the Laguerre polynomials are only exact in the limit $\Delta\rightarrow 0^+$ and both the positions and the number
of level crossings on each level depend on the value of $\Delta$. 

In the following we discuss the exact exceptional solutions of the AQRM to gain a better understanding of the existence of crossing points, 
and the effects of non-zero $\Delta$.

\subsection{Exact exceptional solutions}

The crossing points are known as Juddian points \cite{Judd_1979}, 
which can be calculated through constraint polynomials in terms of the system parameters \cite{Kus1985,Zhong_2014,Li_2015,Li_2016,Kimoto_2020}. 
Importantly, the Juddian points are exactly solvable for arbitrary parameters.

We recall the recurrence relation for the constraint polynomials of the AQRM \cite{Li_2015}
\begin{equation}\label{JuddRecursion}
	\begin{aligned}
		&P_0^n(g,\Delta,\epsilon) = 1,\quad P_1^n(g,\Delta,\epsilon) = 4g^2 + \dfrac{\Delta^2}{4} - \omega^2 - \epsilon \omega, \\
		&P_k^n(g,\Delta,\epsilon) = \left(4 k g^2 + \dfrac{\Delta^2}{4} - k^2\omega^2 - k\epsilon\omega \right) P_{k-1}^n(g,\Delta,\epsilon) \\
		& \phantom{P_k^n(g,\Delta,\epsilon) =}- 4k(k-1)(n-k+1)g^2 P_{k-2}^n(g,\Delta,\epsilon).
	\end{aligned}
\end{equation} 
The polynomials $P^n_n(g,\Delta,\epsilon)$ then determine the degenerate points of the $n$th pair of levels in the AQRM. 

From the constraint polynomials, the number of degenerate points on each level and for each value of $\epsilon$ 
is known \cite{Li_2015,Wakayama_2017,Kimoto_2020}, and consequently all the {necessary} information determining the 
topological properties of the AQRM is known. 
{Moreover, the CI's, determined by the constraint polynomials, are the sources of Berry curvature and thus the geometric phases}.

We define the normalized constraint polynomials
\begin{equation}\label{NormalizedConstraint}
	K_n^\epsilon(g,\Delta) = \dfrac{P_n^n(g,\Delta,\epsilon)}{P_n^n(0,0,0)}.
\end{equation}
A somewhat surprising relation is then given by
\begin{equation}\label{PolynomialsRelation}
	L_n^\epsilon(4g^2) = K_n^\epsilon(g,0). 
\end{equation}
When $\Delta=0$, the constraint polynomials reduce to the corresponding Laguerre polynomials in Eq.~(\ref{AAEigenvalues}), 
which justifies the fact that the AA is only exact in the limit $\Delta\rightarrow 0^+$. 
We demonstrate this correspondence by taking the case $ n=2 $ as an example. 
For $ n=2 $, the normalized constraint polynomial is
\begin{equation}\label{K2}
	K_2^\epsilon(g,\Delta)  = 1-8g^2 + 8g^4 + \dfrac{3\epsilon}{2} - 4g^2\epsilon + \dfrac{\epsilon^2}{2} + \Delta^2 \left(-\dfrac{5}{16} + \dfrac{3g^2}{4} - \dfrac{3\epsilon}{16}\right) + \dfrac{\Delta^4}{64}, 
\end{equation}
and the corresponding Laguerre polynomial is
\begin{equation}\label{L2}
	L_2^\epsilon(4g^2) = 1-8g^2 + 8g^4 + \dfrac{3\epsilon}{2} - 4g^2\epsilon + \dfrac{\epsilon^2}{2}.
\end{equation}
The relation in Eq.~(\ref{PolynomialsRelation}) is readily confirmed with $\Delta=0$. 

We may understand this correspondence with more physical intuition in the displaced oscillator picture. 
In deriving the AA, the small $\Delta/\omega$ term is regarded as a perturbation and the tunnelling processes 
between non-degenerate eigenstates of the displaced oscillators are neglected. 
However, this assumption is never exact since the remote tunnelling is never strictly zero. 
This inaccuracy is even more obvious when $\Delta/\omega$ beomes large. 
With the correspondence demonstrated above, the constraint polynomials can be thought of as the Laguerre polynomials with the corrections coming from extra effects induced by nonzero $\Delta$.

\subsection{Generalized adiabatic approximation}

Given the above observations it is now possible to construct a simple approximation that exactly predicts the CIs in the AQRM. 
This is carried out by replacing the ``topologically-inaccurate" part in the AA with its exact counterpart.
This approach is referred to as the generalized adiabatic approximation (GAA) \cite{Li2021GAA}. 

As a consequence, we have the eigenvalues
\begin{equation}\label{GAAEigenvalues}
	E_{n,\pm}^\text{GAA} = \left(n+\dfrac{l}{2}\right)\omega - \dfrac{g^2}{\omega} \pm \dfrac{1}{2}\sqrt{\left(\Omega_{nl}^\text{GAA}\right)^2 + \left(\epsilon-l\omega\right)^2}, 
\end{equation}
where new tunnelling strengths are defined by 
\begin{equation}\label{GAAoverlap}
	\begin{aligned}
		\Omega_{nl}^\text{GAA} = \Delta \exp\left[{-\dfrac{2g^2}{\omega^2}}\right]\left(-\dfrac{2g}{\omega}\right)^l\sqrt{\dfrac{n!}{(n+l)!}} \, K_n^\epsilon\left(g,\Delta\right). 
	\end{aligned}
\end{equation}
Here $ K_n^\epsilon $ are the normalized constraint polynomials defined in Eq.~(\ref{NormalizedConstraint}).

The eigenstates of the GAA take the same form as those in the AA, namely
\begin{equation}\label{GAAEigenstates}
	\begin{aligned}
		&\psi^\text{GAA}_{n,+} =  \cos\dfrac{\theta_n}{2}| n_+, + \rangle + \sin\dfrac{\theta_n}{2}| (n+l)_-, - \rangle, \\
		&\psi^\text{GAA}_{n,-} =  -\sin\dfrac{\theta_n}{2}| n_+, + \rangle + \cos\dfrac{\theta_n}{2}| (n+l)_-, - \rangle, \\
	\end{aligned}
\end{equation}
with $\theta_n$ now determined by
\begin{equation}\label{GAATheta}
	\tan\theta_n = \dfrac{\Omega_{nl}^\text{GAA}}{\epsilon-l\omega}. 
\end{equation}

The GAA provides a number of important improvements when compared to the AA.
The energy eigenvalues of the AQRM versus coupling strength $ g $ for various values of $\Delta$ and $\epsilon$ are displayed in Fig.~\ref{GAAE}. 
The results determined by exact numerical diagonalization, the AA and the GAA are shown for comparison.
For the calculations by numerical diagonalization, we truncate the dimensions of the Hilbert space when the results converge to precision $\sim 10^{-5}$. 
To calculate the higher levels, setting the number of resonator excitations to be $N=50$ is sufficient.
The adiabatic regime where $\Delta/\omega\ll1$ is displayed in Figs.~\ref{spectra1}-\ref{spectra3}. 
In this regime, both the AA and the GAA are seen to approximate the AQRM with very good agreement. 
Only marginal improvements around level crossings are present. 
Beyond this adiabatic regime, as displayed in Fig.~\ref{spectra4}-\ref{spectra6}, 
the AA substantially deviates from the exact numerical results and unphysical level crossings are induced. 
It should be noted that both the AA and the GAA deviate from the numerical results for small values of $g/\omega$ when $\epsilon$ is nonzero. 
To fix this deviation, higher order corrections need to be taken into account, as done in Refs \cite{Irish2007,Zhang_2013}, where the generalized rotating wave approximation (GRWA) is derived based on the AA. However, the simple form of the energy expression will not be maintained. 
We emphasize that, in the case where $\epsilon$ is nonzero, 
the GAA still outperforms the AA by correctly predicting the level crossings points, and thus the CIs. 
We expect that if the GAA is applied to derive a new version of the GRWA, the deviation for small values of $g/\omega$ will also be fixed.

We conclude that the validity of the AA is reasonable when $\Delta/\omega<1$, the deviation being small when $\Delta/\omega~<0.5$, 
where the AA is at its most powerful. Beyond this value, the deviation becomes noticeable. 
Although the AA reproduces the CIs in the spectrum, the positions of the CIs are never exact. 
This drawback, which has implications for topological properties, is fixed by the GAA. 
The comparison of the energy spectrum shows that the use of the constraint polynomials pushes the validity of the 
perturbative AA into non-perturbative parameter regimes.

\begin{figure}[t]
	\centering
	\subfigure[]{
		\includegraphics[width=.45\linewidth]{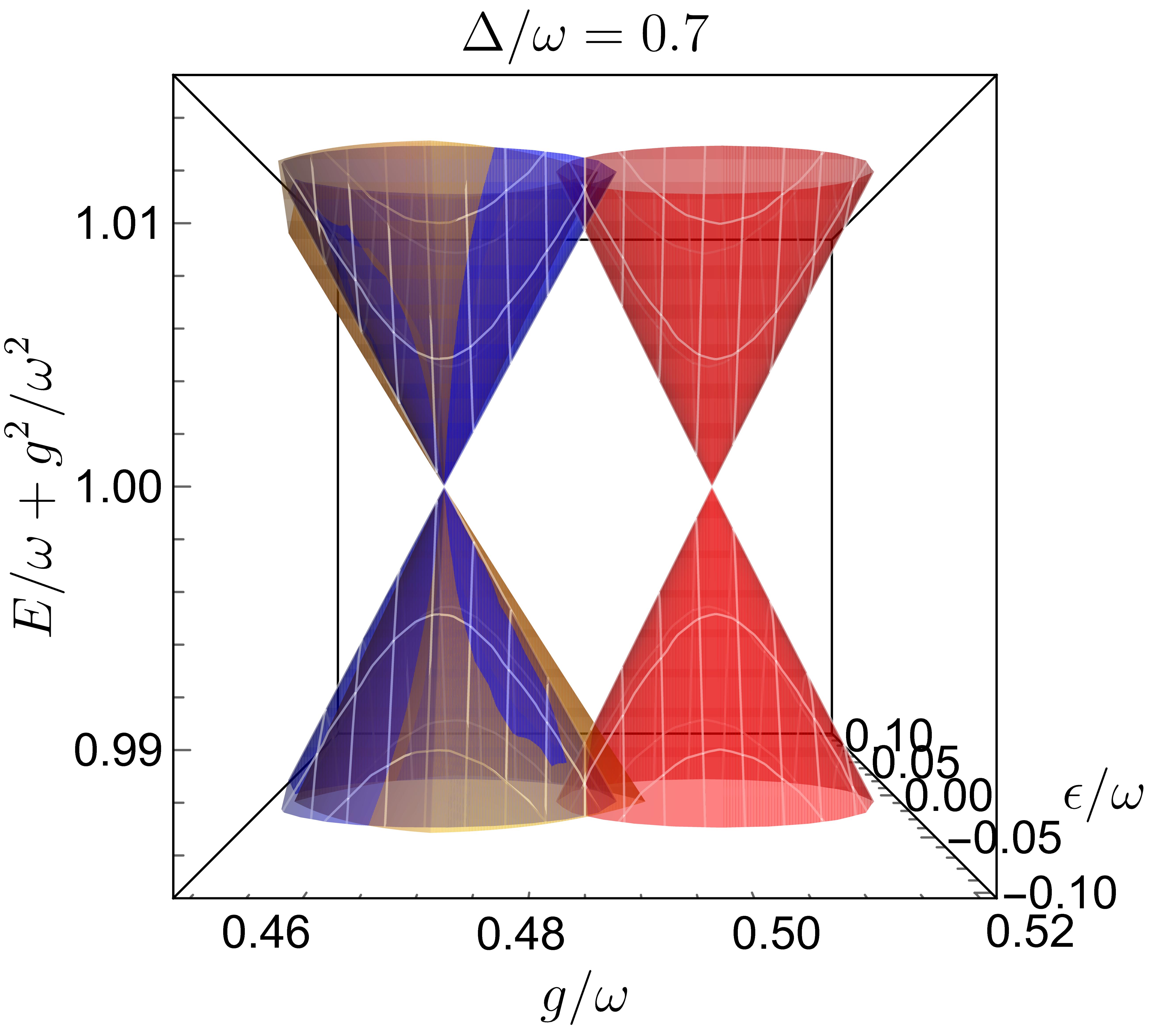}
		\label{GAA3D1}
	}
	\subfigure[]{
		\includegraphics[width=.45\linewidth]{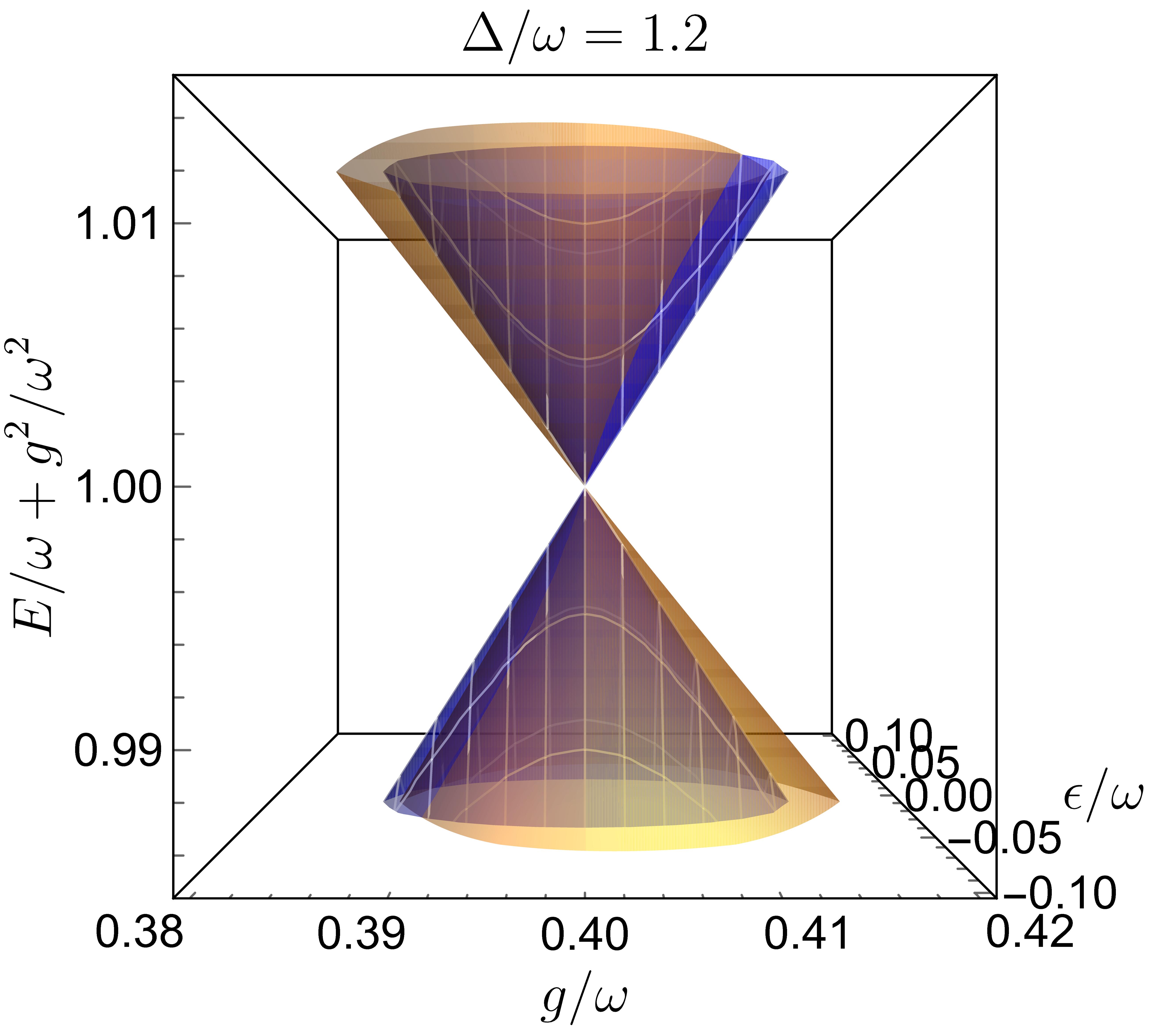}
		\label{GAA3D2}
	}\\
	\subfigure{\includegraphics[width=.3\linewidth]{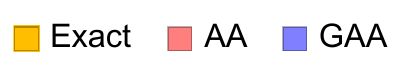}}
	\caption{Lowest conical intersection of the AQRM calculated from exact diagonalization (brown), adiabatic approximation (red) and generalized adiabatic approximation (blue). The parameter values are (a) $\Delta/\omega=0.7$ and (b) $\Delta/\omega=1.2$. }
	\label{GAA3DFig}
\end{figure}

The CIs in the AQRM can be well approximated by the GAA, with the exact intersection points.
The lowest (leading) cone in the energy landscape of the AQRM is displayed in Fig.~\ref{GAA3DFig} with different values of $\Delta/\omega$. 
In Fig.~\ref{GAA3D1} with $\Delta/\omega=0.7$, we observe that both the AA and the GAA reproduce CIs. 
The CI predicted by AA deviates from the exact results while the GAA is accurate in the vicinity of the CI. 
With larger $\Delta/\omega=1.2$ in Fig.~\ref{GAA3D2}, the CI of the AA is out of range whereas the GAA still gives an accurate prediction. 

As discussed so far the GAA has some significant advantages compared to the AA, 
especially in the non-perturbative parameter regimes.
Some limits to its applicability also exist, however.
Values of $\epsilon$ beyond the assumption $ \epsilon - k\omega\ll1 $ 
cannot be easily dealt with in the displaced oscillator. 
For $\epsilon$ values far from the integers, the GAA may therefore break down; 
although there are no CIs for non-integer $\epsilon/\omega$. 
Also, if $\Delta/\omega$ is extremely large, unphysical level crossings predicted by the GAA. 
However, when studying the CIs, we can replace the normalized constraint polynomials $ K_n^\epsilon $ appearing in Eq.~(\ref{GAAoverlap})  
with a ``brute-force" normalization factor. For example, consider 
\begin{equation}\label{Kbar}
	\bar{K}_n^\epsilon = \dfrac{1}{2}\arctan\left[P_n^n(g,\Delta,\epsilon)\right], 
\end{equation}
which shares the same roots as the constraint polynomials and consequently leaves the locations of CIs exact.
The advantage is that Eq.~(\ref{Kbar}) never induces unphysical level crossings for arbitrary parameter values. 
The sacrifice is that the regular eigenvalues are not as accurate as before.

In  conclusion, we note that the AQRM under the GAA can be described by the simple Hamiltonian
\begin{equation}\label{TEH}
	H_n^\text{GAA} = \left(a^\dagger a+\dfrac{l}{2}\right)\omega - \dfrac{g^2}{\omega} + \dfrac{1}{2}(\epsilon-l\omega) \sigma_x^{(n)} + \dfrac{1}{2}K_n^\epsilon  \sigma_z^{(n)}, 
\end{equation}
where $\sigma_{x,z}^{(n)}$ are the same as Eq.~(\ref{AASigmas}). 
This Hamiltonian can be regarded as a simple combination of a biased qubit and a quantum harmonic oscillator. 
By construction, this Hamiltonian shares the CIs of the AQRM and thus can be considered as a topologically equivalent version of the AQRM.

\section{Geometric phases around conical intersections}\label{SectionGP}

As a preliminary example of the applications of the GAA we calculate the geometric phases around CIs associated with the AQRM. 
This is possible because the GAA predicts the locations of CIs exactly.

Suppose the system Hamiltonian $H(\mathbf{R})$ depends on the vector $\mathbf{R} = \left\{R_1, R_2, \dots, R_m \right\}$ of $m$ real parameters.
If the system is initially in an eigenstate, and the parameters $ \mathbf{R} $ are varied slowly enough, 
the system will stay in the  eigenstate corresponding to the Hamiltonian with the instantaneous parameters $ \mathbf{R} $. 
The stationary Schr\"odinger equation is then  
\begin{equation}\label{SSE}
	H(\mathbf{R}) | \psi_n(\mathbf{R}) \rangle = E_n(\mathbf{R}) | \psi_n(\mathbf{R}) \rangle	.
\end{equation}
If we consider the time evolution of the system, the time-dependent Schr\"odinger equation gives rise to a regular dynamical phase that depends on the evolution time period, and a geometric phase that only depends on the path taken in the parameter space. 
For a closed path $ \mathcal{C} $ in parameter space, the general form of the geometric phase is
\begin{equation}\label{SBP}
	\gamma_n = \oint_c \mathbf{A}_n(\mathbf{R})\cdot\mathrm{d}\mathbf{R},
\end{equation}
where the vector potential
\begin{equation}\label{SBerryConnection}
	\mathbf{A}_n(\mathbf{R}) = \mathrm{i} \left< \psi_n(\mathbf{R}) \right| \nabla_\mathbf{R} \left| \psi_n(\mathbf{R}) \right>
\end{equation}
is known as the Berry connection in parameter space.

We now calculate the geometric phases in the AQRM using the eigenstates obtained from the GAA. 
In our case, we consider a two-dimensional parameter space with the vector $\mathbf{R} = \left( g, \epsilon\right)$ since $\Delta$ and $\omega$ are conventionally fixed. 
Since the geometric phase vanishes for real eigenvectors, we need to introduce some imaginary factors in the system \cite{Bohm2003,Mailybaev_2006}. 
From the displaced oscillator basis $\left\{|n_+,+\rangle, |(n+l)_-,-\rangle\right\}$, we construct a new basis
\begin{equation}\label{SBPbasis}
	\begin{aligned}
		&|\phi_n^+\rangle = \dfrac{1}{\sqrt{2}}\left(|n_+,+\rangle + \mathrm{i} \, |(n+l)_-,-\rangle\right), \\
		&|\phi_n^-\rangle = \dfrac{1}{\sqrt{2}}\left(|n_+,+\rangle - \mathrm{i} \, |(n+l)_-,-\rangle\right).
	\end{aligned}
\end{equation}
The $2\times 2$ matrix block Eq.~(\ref{AABlockSigmaForm}) is rewritten in this basis as
\begin{equation}\label{SBPBlock}
	H'_n = 
	\begin{pmatrix}
		h_{11} & h_{12} \\
		h_{21} & h_{22}
	\end{pmatrix},
\end{equation}
with
\begin{equation}
	\begin{aligned}\label{Shs}
		&h_{11} =  h_{22} = \left( n + \dfrac{l}{2}\right)\omega - \dfrac{g^2}{\omega}, \\
		&h_{12} =  h^*_{21} = \dfrac{1}{2}(\epsilon - l \omega) - \dfrac{\mathrm{i}}{2} \Omega_{nl}. 
	\end{aligned}
\end{equation}

The eigenvalues remain unchanged, as in Eq.~(\ref{AAEigenvalues}), whereas the eigenstates are now
\begin{equation}
	\begin{aligned}\label{SBPEigenstates}
		& |\psi_n^+\rangle = \dfrac{1}{\sqrt{2}}\left(|\phi_n^+\rangle + e^{\mathrm{i}\theta_n}|\phi_n^-\rangle\right), \\
		& |\psi_n^-\rangle = \dfrac{1}{\sqrt{2}}\left(-e^{-\mathrm{i}\theta_n}|\phi_n^+\rangle + |\phi_n^-\rangle\right),
	\end{aligned}
\end{equation}
in which $\theta_n$ is again determined through Eq.~(\ref{GAATheta}). 
More generally, the eigenvalues remain unchanged with respect to additional multiplicative factors of $e^{\pm \mathrm{i}\theta_n}$, 
which constitutes a gauge freedom. 
Throughout this work we have employed the gauge given by Eq.~(\ref{SBPEigenstates}).

\begin{figure}[t]
	\centering
	\includegraphics[width=0.7\linewidth]{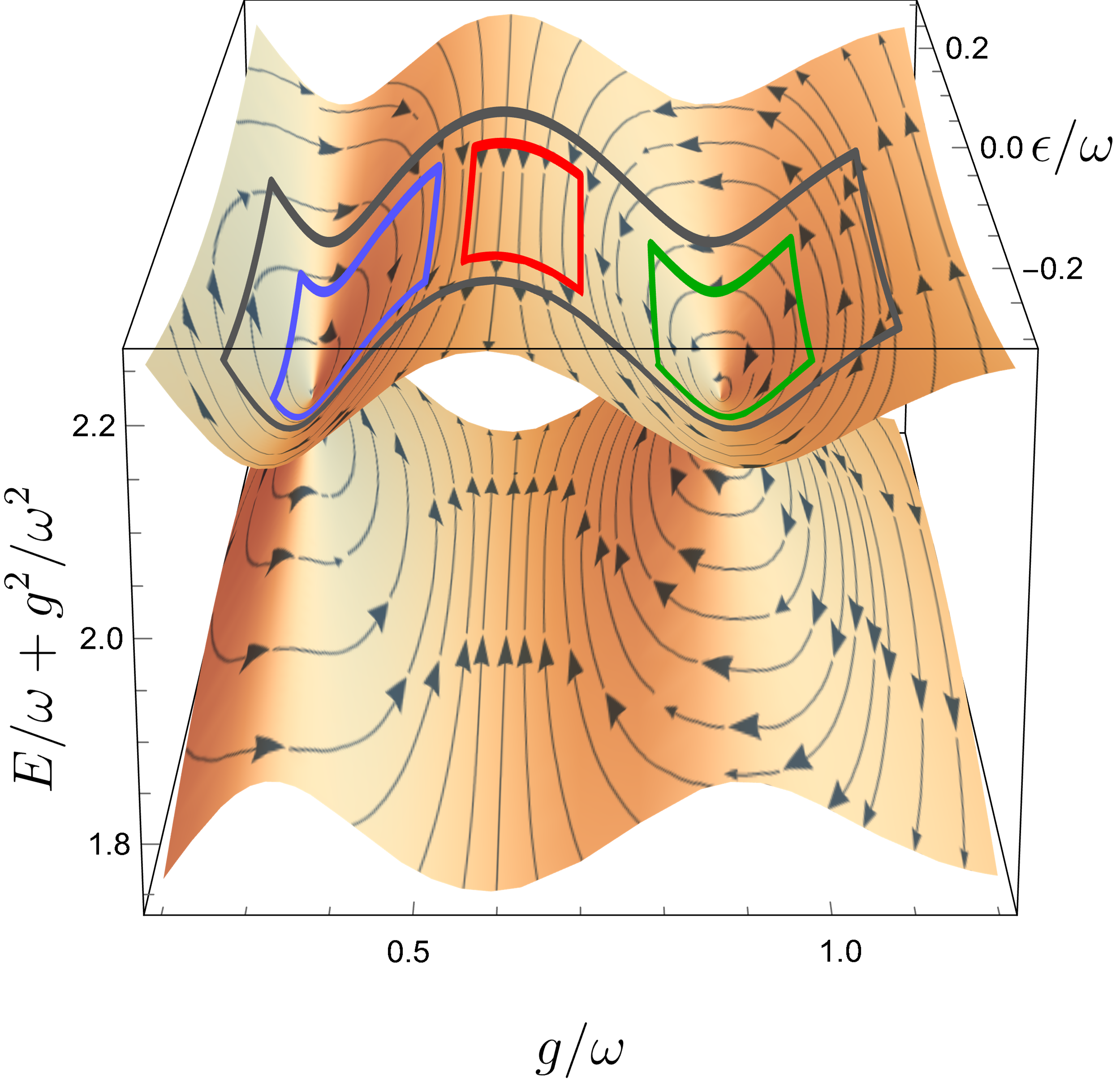}
	\caption{Energy surfaces of the AQRM and the corresponding
		Berry connections under the GAA. The fixed parameters are $\Delta=1$ and $\omega=1$. 
		We consider the four denoted trajectories (blue, red, green and black) in the $(g,\epsilon)$ parameter space. }
	\label{AABP2}
\end{figure}

Having obtained the eigenstates, we can compute the corresponding geometric phases. 
We consider a loop in the two-dimensional parameter space ($g,\epsilon$) with fixed $\Delta$. 
The change in $\theta_n$ must be a multiple of $2\pi$, i.e.,
\begin{equation}\label{Sbetachange}
	\theta_n^f - \theta_n^i = 2m\pi,\quad m\in \mathbb{Z},
\end{equation}
where $\theta_n^i$ and $\theta_n^f$ are the initial and final values of $\theta_n$, respectively. 
The geometric phases for the $n$th pair of states are then calculated as
\begin{equation}\label{BPinAQRM}
	\begin{aligned}
		\gamma_\pm^n =&  \mathrm{i} \oint_c \left< \psi_n^\pm \left| \dfrac{\mathrm{d}\theta_n}{\mathrm{d}\mathbf{R}}\dfrac{\mathrm{d}}{\mathrm{d}\theta_n} \right| \psi_n^\pm \right> \cdot \mathrm{d}\mathbf{R}  = \mp m\pi.\\
	\end{aligned}
\end{equation}
Integer $m$ characterizes the configuration of degenerate points encircled by the parameter loop \cite{Bohm2003}. 
Specifically, the geometric phase is $\pm\pi$ when an odd number of CIs are within the loop and zero when an even number of CIs are circled. 
As a consequence, the wavefunction changes sign when $ m $ takes odd values. 
The geometric phases obtained in Eq.~(\ref{BPinAQRM}) are topological in the sense that they cannot vary smoothly.

Eq.~(\ref{BPinAQRM}) gives the geometric phases associated with CIs in the AQRM. 
Consider, for example, the 6th level of the AQRM. 
{The corresponding energy surfaces, featuring the two CIs},  
are displayed in Fig.~\ref{AABP2}, on which we consider four trajectories.
The blue ($g/\omega\in[0.2,0.5], \epsilon/\omega\in[-0.1,0.1]$) and green ($g/\omega\in[0.8,1], \epsilon/\omega\in[-0.1,0.1]$) 
trajectories each encircle a CI,
whereas the red trajectory ($g/\omega\in[0.55,0.7], \epsilon/\omega\in[-0.1,0.1]$) does not.
The large black loop ($g/\omega\in[0.25,1.1], \epsilon/\omega\in[-0.15,0.15]$) encircles the two CIs. 
Numerical calculations yield the corresponding geometric phases
\begin{equation}\label{BPAA12}
	\gamma^\text{blue} = \pi,\quad \gamma^\text{green} = -\pi, \quad \gamma^\text{black} = \gamma^\text{red} =  0 , 
\end{equation}
each as expected \cite{Berry1984,SW1989}.
{Importantly,} the exact diagonalization of the AQRM in the appropriate basis also gives the same geometric phases and justifies the above derivations. 
{Moreover, the above results remain valid for the AQRM with arbitrary parameter values, which can be verified numerically. 
This is supported by the fact that all crossing points in the AQRM are doubly degenerate. }

\section{Conclusion}\label{SectionConclusions}

In summary, based on the exact solutions for the degenerate crossing points, 
we have proposed a relatively simple generalized adiabatic approximation to the AQRM. 
The GAA maintains the simplicity of the widely adapted  AA and provides substantial improvements to the existing approximations. 
By construction, the GAA predicts the exactly known exceptional solutions. 
It approximates the regular spectrum with very good agreement in large parameter regimes. 
Importantly, the fact that GAA correctly recovers the CIs of the AQRM makes it possible to 
explore the topological properties of the AQRM exactly in analytic fashion. 
We have thus investigated the topological properties around the CIs {in the energy landscape} of the AQRM. 
As to be expected for a model of this kind, the geometric phases are always multiples of $\pi$, 
{with the precise multiple} depending on the configuration of CIs encircled by the trajectories considered.

The GAA approach can be readily applied to other light-matter interaction models, 
such as the anisotropic Rabi model and the Rabi-Stark model, 
where the hidden symmetry is also present \cite{Li2021a,Lu_b} and the Juddian points are exactly solvable. 
An interesting property of these two models is that a parameter-dependent conical intersection occurs in the ground state, 
and their topological transitions could be explored analytically within the framework of the GAA.

Of particular interest, also, is the prospect of experimental measurement and manipulation of the geometric phases of the AQRM within cQED circuits and devices, where often the asymmetric bias term appears naturally. 
In particular, the most {accessible}  cones in the AQRM energy spectrum occur within the deep-strong coupling regime reached in recent experiments \cite{Yoshihara_2016,Yoshihara_2018}, and which are theoretically accessible via the GAA.

\section*{Acknowledgments}

	This work is supported by the Australian Research Council through Discovery Grants DP170104934 and DP180101040. 

\section*{References}

\bibliographystyle{iopart-num}
%\bibliography{../../Bibliography/RabiModelBib}

\begin{thebibliography}{10}
	\expandafter\ifx\csname url\endcsname\relax
	\def\url#1{{\tt #1}}\fi
	\expandafter\ifx\csname urlprefix\endcsname\relax\def\urlprefix{URL }\fi
	\providecommand{\eprint}[2][]{\url{#2}}
	
	\bibitem{Rabi_1936}
	Rabi I~I 1936 {\em Phys. Rev.\/} {\bf 49} 324--328
	
	\bibitem{Rabi_1937}
	Rabi I~I 1937 {\em Phys. Rev.\/} {\bf 51} 652--654
	
	\bibitem{Xie_2017}
	Xie Q, Zhong H, Batchelor M~T and Lee C 2017 {\em J. Phys. A: Math. Theor.\/}
	{\bf 50} 113001
	
	\bibitem{cQEDa}
	Blais A, Girvin S~M  and Oliver W~D 2020 
%	Quantum information processing and quantum optics with circuit quantum electrodynamics. 
	{\em Nat. Phys.\/} {\bf16} 247--256 
%	https://doi.org/10.1038/s41567-020-0806-z

	\bibitem{cQEDb}
	Clerk A~A, Lehnert K~W, Bertet P, Petta J~R and Nakamura Y 2020 
%	Hybrid quantum systems with circuit quantum electrodynamics 
	{\em Nat. Phys.\/} {\bf16} 257--267 
%	https://doi.org/10.1038/s41567-020-0797-9

	\bibitem{cQEDc}
	Carusotto I, Houck A~A, Koll\'ar A J, Roushan P, Schuster D~I and Simon J 2020 
%	Photonic materials in circuit quantum electrodynamics 
	{\em Nat. Phys.\/} {\bf16} 268--279 
%	https://doi.org/10.1038/s41567-020-0815-y
	
	\bibitem{Niemczyk2010}
	Niemczyk T, Deppe F, Huebl H, Menzel E~P, Hocke F, Schwarz M~J, Garcia-Ripoll
	J~J, Zueco D, H\"ummer T, Solano E, Marx A and Gross R 2010 {\em Nat. Phys.\/}
	{\bf 6} 772--776
	
	\bibitem{Yoshihara_2016}
	Yoshihara F, Fuse T, Ashhab S, Kakuyanagi K, Saito S and Semba K 2016 {\em Nat.
		Phys.\/} {\bf 13} 44--47
		
	\bibitem{Frisk2019}
	Kockum A~F, Miranowicz A, Liberato S D, Savasta S and Nori F 2019 
	{\em Nat. Rev. Phys.\/} {\bf 1} 19
	
	\bibitem{Forn2019} Forn-D\'iaz P, Lamata L, Rico E, Kono J and Solano E 2019 
	{\em Rev. Mod. Phys.\/} {\bf 91} 025005

        \bibitem{Blais2020} Blais A, Grimsmo A~L, Girvin S~M and Wallraff A 2021 
 	{\em Rev. Mod. Phys.\/} {\bf 93} 025005
	
	\bibitem{Braak_2011}
	Braak D 2011 {\em Phys. Rev. Lett.\/} {\bf 107} 100401
	
	\bibitem{Chen2012}
	Chen Q~H, Wang C, He S, Liu T and Wang K~L 2012 {\em Phys. Rev. A\/} {\bf
		86} 023822
	
	\bibitem{Zhong_2014}
	Zhong H, Xie Q, Guan X, Batchelor M~T, Gao K and Lee C 2014 {\em J. Phys. A:
		Math. Theor.\/} {\bf 47} 045301
		
	\bibitem{Maciejewski_2014}
	Maciejewski A~J, Przybylska M and Stachowiak T 2014 {\em Phys. Lett. A\/} {\bf
		378} 3445--3451
		
	\bibitem{Li_2015}
	Li Z~M and Batchelor M~T 2015 {\em J. Phys. A: Math. Theor.\/} {\bf 48} 454005
	
	\bibitem{Wakayama_2017}
	Wakayama M 2017 {\em J. Phys. A: Math. Theor.\/} {\bf 50} 174001

	\bibitem{Kimoto_2020}
	Kimoto K, Reyes-Bustos C and Wakayama M 2020 {\em Int. Math. Res. Not.\/} rnaa034
	
	\bibitem{Li2021}
	Li Z~M, Ferri D and Batchelor M~T 2021 {\em Phys. Rev. A\/} {\bf 103} 013711
	
	\bibitem{Zhang_2013}
	Zhang Y~Y, Chen Q~H and Zhao Y 2013 {\em Phys. Rev. A\/} {\bf 87} 033827
	
	\bibitem{Mao_2018}
	Mao B~B, Liu M, Wu W, Li L, Ying Z~J and Luo H~G 2018 {\em Chin. Phys. B\/}
	{\bf 27} 054219
	
	\bibitem{Xie2020}
	Xie W, Mao B~B, Li G, Wang W, Sun C, Wang Y, You W~L and Liu M 2020 {\em J.
		Phys. A: Math. Theor.\/} {\bf 53} 095302
		
	\bibitem{Irish_2005}
	Irish E~K, Gea-Banacloche J, Martin I and Schwab K~C 2005 {\em Phys. Rev. B\/}
	{\bf 72} 195410
		
	\bibitem{Semple_2017}
	Semple J and Kollar M 2017 {\em J. Phys. A: Math. Theor.\/} {\bf 51} 044002
	
	\bibitem{Ashhab_2020}
	Ashhab S 2020 {\em Phys. Rev. A\/} {\bf 101} 023808
	
	\bibitem{Hausinger_2010}
	Hausinger J and Grifoni M 2010 {\em Phys. Rev. A\/} {\bf 82} 062320

	
	\bibitem{Batchelor2016}
	Batchelor M~T, Li Z~M and Zhou H~Q 2016 {\em J. Phys. A: Math. Theor.\/} {\bf
		49} 01LT01
	
	\bibitem{Berry1984}
	Berry M~V 1984 {\em Proc. R. Soc. A\/} {\bf 392} 45--57
	
	\bibitem{SW1989}
	Shapere A and Wilczek F 1989 {\em Geometric Phases in Physics\/} (World Scientific, Singapore)
	
	\bibitem{Bohm2003}
         B\"ohm A, Mostafazadeh A, Koizumi H, Niu Q and Zwanziger J 2003 {\em The Geometric Phase in Quantum Systems\/}
         (Springer Berlin Heidelberg)
	
	\bibitem{Cohen_2019}
	Cohen E, Larocque H, Bouchard F, Nejadsattari F, Gefen Y and Karimi E 2019
	{\em Nature Reviews Physics\/} {\bf 1} 437 
	
	\bibitem{Larson2020} 
	Larson J, Sj\"oqvist E and \"Ohberg P 2020 {\em Conical Intersections in Physics} (Springer-Verlag GmbH)
	
	\bibitem{Leek_2007} Leek P~J, Fink J~M, Blais A, Bianchetti R, Goppl M, Gambetta J~M, Schuster D~I, Frunzio L, 
	Schoelkopf R~J and Wallraff A 2007
	{\em Science\/} {\bf 318} 1889--1892
	
	\bibitem{Gasparinetti_2016} Gasparinetti S, Berger S, Abdumalikov A~A, Pechal M, Filipp S, and Wallraff A 2016 
	{\em Science Advances} {\bf 2} e1501732 
	
	\bibitem{Mottonen_2008} M\"ott\"onen M, Vartiainen J~J and Pekola J~P 2008 
	{\em Phys. Rev. Lett.\/} {\bf 100} 177201 
	
	\bibitem{Berger_2012} Berger S, Pechal M, Pugnetti S, Abdumalikov A~A, Steffen L, Fedorov A, Wallraff A and Filipp S 2012
	{\em Phys. Rev. B\/} {\bf 85} 220502 
	
	\bibitem{Fuentes_Guridi_2002} Fuentes-Guridi I, Carollo A, Bose S and Vedral V 2002
         {\em Phys. Rev. Lett.\/} {\bf 89} 220404 
	
	\bibitem{Larson_2012} Larson J 2012 {\em Phys. Rev. Lett.} {\bf 108} 033601 
	
         \bibitem{Deng_2013} Deng W-W and Li G-X 2013 {\em J. Phys. B: At., Mol. Opt. Phys.\/} {\bf 46} 224018 
         
         
         \bibitem{Wang_2015}  Wang M, Wei L  and Liang J 2015 {\em Phys. Lett. A \/} {\bf 379} 1087-1090
         
         \bibitem{Mao_2015} Mao L, Huai S, Guo L and Zhang Y 2015 
         {\em Ann. Phys.\/} {\bf 362} 538 
         
         \bibitem{Calder_n_2016} Calder\'on J and De Zela F 2016 
         {\em Phys. Rev. A\/} {\bf 93} 033823 
         
         \bibitem{Wang2019} Wang Y and Luo X 2019 {\em Opt. Commun.\/} {\bf 451} 13-16
         
         \bibitem{HLH_1963} Herzberg G and Longuet-Higgins H~C 1963
         {\em Disc. Faraday Soc.\/} {\bf 35} 77--82
		
	\bibitem{Li2021GAA}
	Li Z~M and Batchelor M~T  \textit{Generalized adiabatic approximation to the quantum Rabi model} \eprint{2104.13062}
	
	\bibitem{Braak_2019}
	Braak D 2019 {\em Symmetry\/} {\bf 11} 1259
		
	\bibitem{Mangazeev_2021}
	Mangazeev V~V, Batchelor M~T and Bazhanov V~V 2021 {\em J. Phys. A: Math.
		Theor.\/} {\bf 54} 12LT01
		
	\bibitem{RBW2021} Reyes-Bustos C, Braak D and Wakayama M 2021 
	{\em J. Phys. A: Math. Theor.\/} {\bf 54} 285202
	
	\bibitem{Li2021a}
	Li Z~M and Batchelor M~T 2021 {\em Phys. Rev. A\/} {\bf 103} 023719
	
	\bibitem{Lu_a}
	Lu X, Li Z~M, Mangazeev V~V and Batchelor M~T 2021 
	 {\em J. Phys. A: Math. Theor.\/} {\bf 54} 325202

	\bibitem{Lu_b}
	Lu X, Li Z~M, Mangazeev V~V and Batchelor M~T 
	\textit{Hidden symmetry operators for asymmetric generalized quantum Rabi models} \eprint{2104.14164}
	
	\bibitem{Philbin_2014}
	Philbin T~G 2014 {\em Am. J. Phys\/} {\bf 82} 742--748
	
	\bibitem{RW2021} Reyes-Bustos C and Wakayama M 
	\textit{Degeneracy and hidden symmetry -- an asymmetric quantum Rabi model with an integer bias} 
	\eprint{2106.08916 }	
	
	\bibitem{Judd_1979}
	Judd B~R 1979 {\em J. Phys. C: Solid State Phys.\/} {\bf 12} 1685--1692
	
	\bibitem{Kus1985}
	Ku{\'{s}} M 1985 {\em J. Math. Phys.\/} {\bf 26} 2792--2795
	
	\bibitem{Li_2016}
	Li Z~M and Batchelor M~T 2016 {\em J. Phys. A: Math. Theor.\/} {\bf 49} 369401
	
	\bibitem{Irish2007} Irish E K 2007 {\em Phys. Rev. Lett.\/} {\bf 99} 173601 
			
	\bibitem{Mailybaev_2006}
	Mailybaev A~A, Kirillov O~N and Seyranian A~P 2006 {\em Dokl. Math.\/} {\bf 73}
	129--133
		
	\bibitem{Yoshihara_2018}
	Yoshihara F, Fuse T, Ao Z, Ashhab S, Kakuyanagi K, Saito S, Aoki T, Koshino K
	and Semba K 2018 {\em Phys. Rev. Lett.\/} {\bf 120} 183601
	
	
\end{thebibliography}

\providecommand{\newblock}{}

\end{document}